\documentclass[
    ,final            
  ,numberedheadings 
  ]
  {aipproc}

\layoutstyle{6x9}

\begin{document}

\title{Lighthouses with two lights:  burst oscillations from the
  accretion-powered millisecond pulsars}

\classification{95.85.Nv, 97.10.Kc, 97.60.Gb, 97.60.Jd, 97.80.Jp}
\keywords      {binaries: general, stars: neutron, stars: rotation, X-rays:bursts}

\author{Anna L. Watts}{
  address={Astronomical Institute ``Anton Pannekoek'', University of Amsterdam, Kruislaan 403,\\
    1098 SJ Amsterdam, the Netherlands; A.L.Watts@uva.nl}
}

\begin{abstract}
The key contribution of the discovery of nuclear-powered pulsations
from the accretion-powered
millisecond pulsars (AMPs) has been the establishment of burst
oscillation frequency as a reliable proxy for stellar spin rate.  This
has doubled the sample of rapidly-rotating accreting neutron stars and
revealed the unexpected absence of any stars rotating near the break-up
limit. The resulting `braking problem' is now a major concern for
theorists, particularly given the possible role of gravitational wave
emission in limiting spin.  This, however, is not the only area where
burst oscillations from the AMPs are having an impact.  Burst
oscillation timing is developing into a promising technique for
verifying the level of spin variability in the AMPs (a topic of
considerable debate).  These sources also provide unique input to our
efforts to understand the still-elusive burst oscillation
mechanism. This is because they are the only stars where we can reliably gauge the role of
uneven fuel deposition and, of course, the magnetic field.
\end{abstract}

\maketitle

\section{Introduction}

`History', as George Orwell once noted, `is written by the
winners' \citep{orw44} - or, in this case, by the workshop hosts.  When we gathered in Amsterdam in
April 2008, it was ostensibly to celebrate ten years since the
discovery of the first Accreting Millisecond X-ray Pulsar (AMXP).  We were,
however, a full two years too late.  For the first AMXP was not SAX
J1808.4-3658 \citep{wij98}, but rather the far less well-known 4U
1728-34 \citep{str96b}.  How on earth, you might ask, could such a slip
go unnoticed?  The trick, of course, lies in the terminology.  Most
astronomers (the author included) tend to think of the AMXPs as
comprising only the {\it accretion-powered} millisecond pulsars (AMPs),
forgetting the equally large class of {\it nuclear-powered} millisecond
pulsars (NMPs) - the burst oscillation sources.  

Most of this volume focuses on the AMPs, where persistent pulsations
are generated as accreting
material is channeled by the magnetic field onto magnetic polar caps
that are offset from the rotational poles.  The NMPs, by contrast,
show pulsations during Type I X-ray bursts (thermonuclear
explosions on the stellar surface caused by rapid unstable
burning of accreted material). The cause of the brightness asymmetry
in the NMPs remains an open question \citep{str06, gal08}, and to do
full justice to NMP phenomenology would merit a much longer
discussion.  In this article, however,
I will focus on the small set of NMPs that are also
AMPs.  These rare objects provide a unique insight into many current
problems in neutron star astrophysics because, as suggested by my
title, they are lighthouses with two different light sources.  The
accretion-powered pulsations tell us how the material arrives on the
stellar surface, while the nuclear-powered pulsations tell us what
happens once it gets there.  

Section \ref{data} provides a brief overview of the relevant
observational results.  The bulk of the review focuses, however, on the
astrophysical questions where these sources have made or are making
a major contribution to our understanding.  These include the spin
distribution of AMXPs, torque modeling, and the burst oscillation
mechanism. 
 
\section{Observational summary}
\label{data}

Type I X-ray bursts have been observed from two of the seven
persistent\footnote{Here I am referring to the persistence
  of accretion-powered pulsations throughout an accretion episode.
  The intermittent pulsators, by contrast, have detectable accretion-powered
  pulsations only sporadically during accretion episodes. Confusingly,
  the term
  persistent is also used for systems that accrete at detectable levels
  (measured via X-ray emission) all the time.  Transients, on the
  other hand, accrete only occasionally, with alternating periods of
  outburst (not to be confused with bursts!) and quiescence.}, and all
three of the intermittent, AMPs. The AMPs for which no bursts have been
detected are
the four ultra-compact systems (XTE J1751-305, XTE J1807-294, XTE
J0929-314 and SWIFT J1756.9-2508) and IGR J00291+5934.  The latter has
a similar orbital period to the bursting AMPs, so is a
good candidate 
for burst detection during its next outburst\footnote{A new outburst
  started just as this article was being completed \citep{cha08}.}.    

The first AMP to be detected as an NMP was
SAX J1808.4-3658 \citep{cha03}. The burst oscillations exhibit
frequency drifts of a few Hz in the rising phase of the brightest
bursts, settling down to a frequency that is within 0.1 Hz of the spin
frequency in the burst tail. This result was followed by the discovery
of burst oscillations in XTE J1814-338 \citep{str03}.
In this source, the nuclear-powered pulsation frequency is extremely
stable, and equal 
to the spin frequency inferred from the accretion-powered pulsations
\citep{wat05}.  

Of the three intermittent AMPs only Aql X-1 has burst
oscillations.  Indeed this source was discovered to be an NMP
\citep{zha98} long before its detection as an intermittent AMP
\citep{cas08}. The frequency inferred from the accretion-powered
pulsations is offset by a small amount ($<$ 1 Hz) from the asymptotic frequency
of the burst oscillations (as in many other sources, the burst
oscillations from Aql X-1 drift in frequency during a burst but
asymptote to a frequency that remains stable from burst to burst \cite{mun02a}).  Burst oscillations have not been 
detected in any of the bursts from HETE J1900.1-2455 or SAX
J1748.9-2021\footnote{The statistical significance of the burst
  oscillation claimed by \cite{kaa03} is much lower
  than quoted \citep{alt08}.} that have been recorded with high time resolution
instruments \citep{gal08}.

\section{Neutron star spin}

\subsection{Spin distribution and the paucity of rapid rotators}

Identifying the spin distribution of the various classes of neutron
star has been a major research goal for many years.  Spin rates
are important to our understanding of stellar and binary evolution:
finding rapidly-rotating accreting neutron stars has been critical,
for example, to confirming the recycling scenario for the formation of the
millisecond radio pulsars \citep{bhat91}.  Maximum spin rates are also
important to relativistic nuclear astrophysics, since they place
firm constraints on equations of state \citep{lat07}.  

Identifying the most rapidly-rotating X-ray pulsars, however, has
proved to be far more difficult than for the radio pulsars.  The sources are far
less numerous, and the vast majority of neutron stars in Low Mass
X-ray Binaries (LMXBs) do not have accretion-powered pulsations.  The
discovery of the NMPs promised a substantial enlargement of the sample
of sources with measured spins, if it could be proven that burst oscillation frequency
was the spin frequency.  Frequency drifts during bursts
\citep{mun02a}, and the failure to identify the mechanism 
responsible for the nuclear-powered pulsations, had led to
some caution - although repeated measurements of the same burst
oscillation frequency for a given source \citep{mun02a}, and the presence of
oscillations at the same frequencies in both Type I bursts and
superbursts \citep{str02}, certainly supported the
hypothesis.  The eventual detection of burst oscillations from the AMPs, at the spin 
frequency, now seems to have resolved this
question.  The correspondence between the two frequencies is not exact
for SAX J1808.4-3658 and Aql X-1 (which have drifting burst
oscillation frequencies), but identification to
within a few Hz is close enough for most practical purposes.   

\begin{figure}[!hpb]
 \includegraphics[height=.3\textheight, clip]{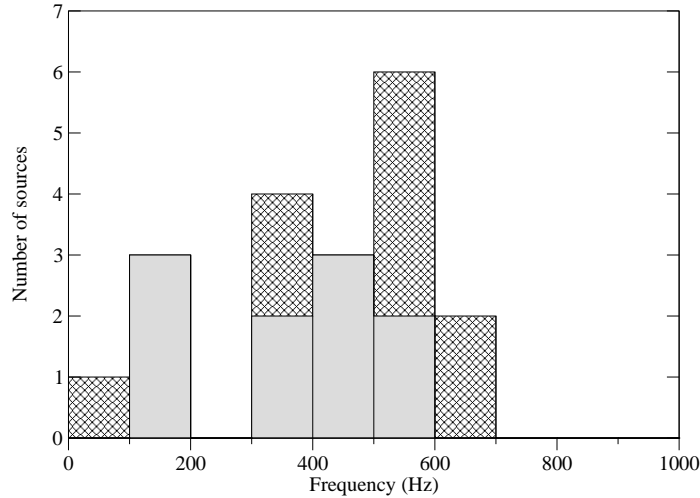}
  \caption{The distribution of measured spin rates inferred from
    the AMPs (grey) and NMPs (hatched).  SAX J1808.4-3658, XTE
    J1814-338 and Aql X-1 are included in the AMP sample, rather than the NMP
    sample.  We only include NMPs where the
    same burst oscillation frequency has been seen in more than one
    burst, since these are the statistically secure detections (see
    \cite{wat08a} for further discussion of this issue). We do not
    include any source with spin below 40 Hz in this Figure.}
\label{spindist}
\end{figure}

The main consequence has been a doubling of the number of
rapidly-rotating accreting neutron stars with a measured spin rate
(Figure \ref{spindist}).  This has brought
to light a new and interesting problem.  Simple estimates of accretion-induced spin-up over the
lifetime of an LMXB suggest that there should be a population of
neutron stars
with spin frequencies above 1 kHz \citep{cook94} (and such rapid spin rates are
permitted by all modern equations of state \cite{lat07}).  However the fastest AMXP
spins at 620 Hz 
\citep{har03}, while the fastest radio
pulsar spins at 716 Hz \citep{hes06}. If the evolutionary estimates
are sound, there is a requirement for a braking mechanism to halt the
spin-up.   Magnetic braking (due to interaction between the stellar
field and the accretion disk) is one possibility \citep{gho78, whi97, and05}. Another, which
has generated a lot of excitement, is the
emission of gravitational waves \citep{papa78, wag84, bil98}.  This could make AMXPs promising
sources for future gravitational wave detectors, although ironically
this is one application where knowing the spin to a very high
degree of precision would be important \citep{wat08a}.

It is of course worth remembering that we still do
not understand the mechanism responsible for burst oscillations.  The
burst oscillation properties of SAX J1808.4-3658 and XTE J1814-338 are rather
unusual compared to the rest of the NMPs (Section \ref{bosc}), and
it is still possible 
that the cause of the nuclear-powered pulsations may differ for these
sources.  The results from  
Aql X-1 are reassuring in 
this regard, but to have full confidence in the use of
burst oscillations 
as a spin proxy the mechanism must be uncovered.

\subsection{Spin variation and torques in neutron stars}

The spin histories of the AMXPs are valuable because of
what they reveal about the properties of the star and the different
torque mechanisms that may operate.  In terms of stellar properties we
are interested in the moments of inertia of, and degree of coupling
between, the various components - particularly the solid crust and superfluid
core \citep{lat07}. Possible angular momentum sources and sinks include
material/magnetic 
torques as matter from the companion accretes via a
magnetically-threaded accretion disk \citep{gho78}, magnetic dipole
radiation, jets \citep{mig06},
and gravitational wave emission (with various different mechanisms
capable of generating a quadrupole \citep{wat08a}).  The challenges
inherent in this type of analysis and 
modeling are well illustrated by earlier studies of the high magnetic field
(slower-rotating) X-ray pulsars \citep{bil97}.  In 
the AMXPs magnetic fields are weaker, but gravitational wave
torques could play a larger role since gravitational wave emission
scales strongly with 
rotation rate. 

Spin histories for the AMXPs are constructed using standard radio
pulsar timing techniques that involve measuring phase shifts between folded
pulse profiles.  Pulse profiles in the AMXPs are however notoriously
variable, since they are affected by fluctuations in the accretion
flow.  This additional noise leads to pulse phase wander which is
sufficiently large 
that it may mimic or mask genuine spin variation.  As evidenced by
the lively discussions on this topic at the Amsterdam workshop, the
level of confidence in inferred values of spin
derivatives remains a matter of vigorous debate \citep{gal02, bur06,
  bur07, pap07, pap08, har08, rig08}. 

Timing using burst oscillations could in principle provide an
independent test of spin variation, since the emission mechanism for the
nuclear-powered pulsations is thought to be quite
different to the accretion-powered pulsations. Drifts in burst
oscillation frequency, of course, complicate this task (making
analysis for SAX J1808.4-3658 very difficult).  However XTE
J1814-338, which has exceptionally stable burst oscillations and a
large burst sample across its one recorded 50 day outburst, is a
highly promising source for nuclear-powered pulsation timing.  Its
accretion-powered pulsations show substantial phase
wander, some of which could be due to changes in
spin frequency \citep{pap07}.  A full timing analysis of the burst
oscillations has now been completed \citep{wat08b}, and the results have proved
surprising.  The nuclear-powered pulsations are completely
phase-locked (with zero offset) to the accretion-powered pulsations,
tracking perfectly all of the phase wander throughout the main part of
the outburst.  At first glance, this does
little to resolve the debate, since the result is certainly consistent
with some or all of the 
changes being caused by spin variation.  What it does do, though, is
to provide a major constraint on models that attempt explain the phase wander
without requiring spin changes.  Any model must now be able
to explain locked jitter in two totally different types of pulsation.
There are mechanisms that could in principle do this (see Section
\ref{bosc} and the discussion in 
\cite{wat08b}), but these need to be tested further.  If they do not
prove viable,  this would support there being at least some genuine spin
variation.

\section{Burst oscillation mechanisms}
\label{bosc}

The nature of the brightness asymmetry that causes nuclear-powered
pulsations is still not understood for any source, AMP or otherwise.
The status of the various different models may be summarized as
follows:  

\begin{itemize}
\item{{\bf Hotspot spread}: A localized growing hotspot is expected
    to exist in the burst rise, since ignition should not
    start simultaneously across the stellar surface \citep{sha82}.
    There is evidence that expanding hot spots are related to the
    presence of burst oscillations in the rising phases of some bursts
    (see for example \cite{str97, str98}).  However only a very small
    percentage of the burst sample has been subject to this rigorous level of
    analysis, and it remains to be seen whether the entire sample is
    consistent with this model.  Where the hotspot model runs into
    real difficulty is in the burst tail.  Once the flame front has spread
    across the star, the 
    remaining asymmetry should not be strong enough to explain the continued
    presence of pulsations \citep{str06}.}
\item{{\bf Thermonuclear hurricanes}: The
    Coriolis force can act to 
    confine the burning area during the burst rise \citep{spi02} (making this model to
    some degree a variant of the spreading hotspot).  This may
    explain the presence of oscillations in the burst rise, although
 like the previous model it has yet to be put to rigorous
    test.  The authors of this study \citep{spi02} conjectured that
    the resulting
    unstable flows might lead to the development of similar localized
    phenomena in the burst tail.  Subsequent simulations
    have apparently not borne this out, although magnetic effects during vortex
    formation remain to be investigated fully (contribution by Levin,
    this workshop). } 
\item{{\bf Surface modes}:  Global oscillations may be excited
    by flame spread, 
    generating a brightness asymmetry that could persist throughout
    the burst tail.  Attention has focused on
    buoyant r-modes in the neutron star ocean, since these have
    frequencies close to the stellar spin rate \citep{hey04}.
    The mode model also provides a natural explanation for frequency drift
    seen in the tails of many bursts, since the frequency will change
    as the surface layers cool.  Unfortunately the
    model over-predicts the observed drifts: coupling to a crustal
    interface wave, which was proposed as a way of limiting the drift
    \citep{pir05b}, has now been shown to be inefficient
    \citep{ber08}.  Alternative mode types including photospheric
    \citep{hey04} and shearing oscillations \citep{cum05} also have
    shortcomings \citep{ber08}, but magnetic effects may play a
    significant role in determining mode behaviour
    (contribution by Cumming, this workshop).}
\end{itemize}

The most significant contribution that the AMPs have made to our
understanding of the burst oscillation mechanism is of course the
requirement that the frequency should lie within a
few Hz of the spin rate: all of the models 
listed above take this as a basic premise.  However these sources are also
valuable in other ways.  They are the only systems in which we can
reliably gauge the role of the magnetic field in generating and
maintaining burst oscillations.  They also let us assess
the influence of asymmetric fuel deposition.  For the weak magnetic
fields of the AMXPs fuel should spread before reaching ignition depth
\citep{bro98} but there are other 
local effects at the deposition point, such as higher
temperature, that may be important. 

These factors have motivated efforts to measure the properties of the
burst oscillations of the AMPs and compare them to both the
accretion-powered pulsations and the other burst oscillation sources.
Key results, drawn from many different references (non-AMPs \cite{mun02a,
  mun02b, mun03, mun04, gal08}; intermittent AMPs \cite{gal07,
  alt08, cas08}; persistent AMPs \cite{cha03, str03, wat05, wat06,
  wat08b, har08b, har08c}) are as follows:  

\begin{enumerate}
\item{For most NMPs, including Aql X-1, burst oscillations are only
    detectable in some 
    bursts.  They are more prevalent at high accretion rates and in short,
    He-rich bursts.  SAX J1808.4-3658 and XTE J1814-338
    show them in every burst despite low accretion rates and, for XTE
    J1814-338, long mixed H/He bursts.}
\item{The amplitudes of burst oscillations in the non-AMPs lie in the
    range 2-20\% RMS.  The absolute amplitudes of the burst oscillations from
    the AMPs
    are similar, but never exceed the accretion-powered pulsation amplitudes.}
\item{SAX J1808.4-3658 and XTE J1814-338 are the only sources where the burst
    oscillations have 
    detectable harmonic content, albeit at a lower amplitude than in the
    accretion-powered pulsations.}
\item{The burst oscillations for most NMPs, and Aql X-1, have an
    amplitude that rises with energy.  For XTE J1814-338, and most of
    the bursts from SAX J1808.4-3658, amplitude falls with 
    energy. This behaviour is inconsistent with both simple hotspot and surface
    mode models \citep{pir06}.  A fall in amplitude with energy is
    also seen in the accretion-powered pulsations of all of the
    persistent AMPs, but it is not clear that the same
    mechanism could explain this effect in both pulsation types.} 
\item{Burst oscillations from the AMPs show no detectable phase lags.
    This behaviour is similar to that of the other NMPs, which show at
    most marginal hard lags.}
\item{The frequency drifts seen in Aql X-1 are typical of the other
    NMPs:  a slow rise during the burst to a saturation frequency.
    The two persistent AMPs with burst oscillations are very 
    different.  For XTE J1814-338 there are no detectable drifts
    except in the final brightest burst, which has a small drop in
    frequency in the burst rise.  For SAX J1808.4-3658 burst
    oscillation frequency rises rapidly by several Hz in the rising
    phase of the bright bursts, in some cases overshooting the
    spin frequency.  Drifts in the burst tails, however, are minimal.}  
\item{The nuclear-powered pulsations in XTE J1814-338 are completely phase-locked to
    the accretion-powered pulsations even though the latter show
    substantial phase wander over the course of the outburst.}
\end{enumerate} 

So what do these results tell us about the burst oscillation
mechanism?  Firstly, that existing models are
no better at explaining AMP burst oscillation properties than they are
at explaining the rest of the NMP population.  The fall of amplitude
with energy, and frequency overshoot, for example, are not predicted by any
current model. 

The second important question is whether we are looking
at a continuum of behaviour that could be attributed to one mechanism
(with differences being set by varying magnetic field strength, for
example).  The burst oscillation properties of
Aql X-1 sit comfortably within the general population, perhaps not
surprisingly for a source that is a very intermittent AMP.
The non-detection of burst oscillations from the other two
intermittent AMPs is also not too worrying, since many LMXBs with
bursts fail to show burst oscillations. SAX J1808.4-3658 is more of a
challenge.  Whether this source is consistent with a unified model
seems to depend primarily on the mechanism
responsible for the 
frequency shift.  The source could fit if, as suggested by
\cite{cha03}, the rapidity and magnitude of the frequency shift could
be shown to depend
strongly on magnetic field strength (or the degree 
of misalignment between the magnetic field and rotational pole).  

The remaining source, XTE J1814-338, is an oddball. The properties of its
nuclear-powered pulsations differ in almost every way from the rest of
the sample.  The phase-locking of the two types of pulsation, however,
suggests that the presence of nuclear-powered pulsations in this source may be
related to premature ignition and subsequent stalling of the flame
front \citep{wat08b}.  This is a 
rather exciting possibility, but is sadly unlikely to explain the
presence of oscillations in the bright, He-rich bursts of the other
sources (flame front stalling being less likely in such bursts).  This leaves open
the intriguing possibility that there may be at least two different burst oscillation mechanisms.

\section{Conclusions}
\label{conc}

The discovery of nuclear-powered pulsations from the AMPs has cemented the
link between burst oscillation frequency and spin frequency. In
addition to confirming the absence of rapid rotators (now a major
problem for evolutionary models), this has imposed the
strongest single constraint on candidate burst oscillation
mechanisms. Despite this huge clue, the mechanism remains elusive:
but continuing analysis of the AMPs is providing tantalising evidence
that is driving the development of new theoretical models.

\bibliographystyle{aipproc}   
\bibliography{annawatts}

\end{document}